\documentclass[fleqn,twoside]{article}
\usepackage{espcrc2}

\usepackage{graphicx}
\usepackage[figuresright]{rotating}

\newcommand{\AmS}{{\protect\the\textfont2
  A\kern-.1667em\lower.5ex\hbox{M}\kern-.125emS}}

\def\mdmatm{\Delta m^2_{32}}
\def\dmatm{$\mdmatm$}

\def\numunue{$\nu_\mu \rightarrow \nu_e$}
\def\anumunue{$\bar\nu_\mu \rightarrow \bar\nu_e$}

\title{Status of the MINOS experiment}

\author{Milind V. Diwan\address{Brookhaven National Laboratory, \\ 
        Upton, NY, 11973}
\thanks{Representing the MINOS collaboration which has $\sim$ 200 
collaborators from 30 institutions.} }

\begin{document}

\begin{abstract}
I will present the status of the long baseline neutrino 
oscillation experiment MINOS at  Fermi National 
Accelerator Laboratory (Fermilab). 
I will summarize the status of the detector and beam construction,  
the expected event rates and sensitivity to physics. I will also
 comment on possible future plans to 
improve the performance of the experiment.  
\vspace{1pc}
\end{abstract}

\maketitle

\section{Introduction}

  The strongest evidence for neutrino
  oscillations comes from astrophysical observations of
  atmospheric neutrinos with $\Delta m^2_{atm} = 
(1.6 - 4.0) \times 10^{-3}
  ~eV^2$ and maximal mixing \cite{sk1,sktau},  and from 
  solar neutrinos with $\Delta m^2_{solar} = (3
  -10) \times 10^{-5} ~eV^2$ assuming the 
LMA solution~\cite{sno1,sno2,sno3}.
  The observation by the LSND
  experiment~\cite{lsnd} will soon be re-tested at Fermilab by the
  mini-Boone \cite{boon} experiment. Therefore we will not discuss it 
  further in this document.
   There is now a consensus that there are four main goals
  in the field of neutrino oscillations 
   that should be addressed soon with accelerator neutrino
  beams:
  \begin{enumerate}
  \item Precise determination of $\Delta m^2_{32}=\Delta m^2_{atm}$
 and $\sin^2 2 \theta_{23}$ 
and definitive observation of oscillatory behavior.
  \item Detection of \numunue{} in the appearance mode.  If the measured
    $\Delta m^2$ for this measurement is near \dmatm{} then this
    appearance signal will show that $\left|U_{e3}\right|^2 (=
    \sin^2\theta_{13})$ from the neutrino mixing matrix in the standard
    parameterization is non-zero.
  \item Detection of the matter enhancement effect in \numunue{} in the
    appearance mode.  This
 effect will also allow us to measure the sign
    of \dmatm{}, i.e. which neutrino is heavier.
  \item Detection of CP violation in neutrino physics.  The neutrino
    CP-violation in the standard neutrino mixing model  
comes from the phase
    of $U_{e3}$ in the mixing matrix. This phase
    causes an asymmetry in the oscillation rates
    \numunue{} versus \anumunue{}.
  \end{enumerate}
In this paper I will describe how the MINOS experiment 
\cite{mprop,nim1,nim2,nim3,nim4} will make the first important 
contributions  to this program of physics.

\subsection{The neutrino beam}

\begin{figure}[htb]
\vspace{9pt}
\includegraphics[width=17pc]{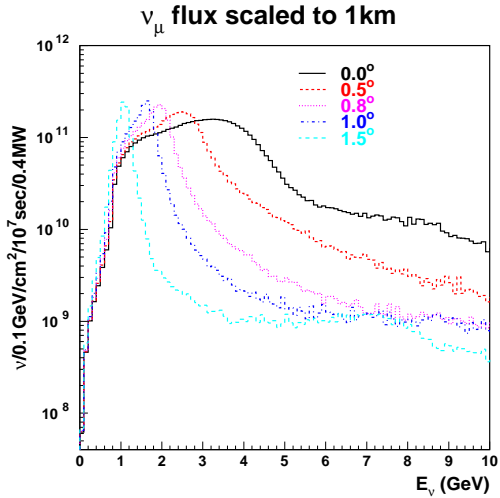}

\caption{The muon neutrino flux expected from the NuMI beam. 
The flux is normalized for 0.4 MW of 120 GeV proton for 
$10^{7}$ sec of running time.
The flux is scaled to be at 1 km distance from 
the NuMI target, but the spectrum shape was calculated  for 735 km.
The spectrum expected at several angles to the $0^o$ axis is
also shown.  }
\label{flux}
\end{figure}

A new neutrino beamline, called NuMI, has been constructed at 
Fermilab \cite{nim3}. High energy, 120 GeV, protons from the 
main injector with intensity of $4\times 10^{13}$ every 
1.9 sec in 8.1$\mu$s-long  pulse will strike a 
1 m-long segmented graphite target. The secondary mesons 
are collected by a two-horn focusing system and directed 
towards the Soudan Iron Mine in Minnesota, 735 km away, 
at an angle of 58 mrad into the ground. 
The mesons, most of which are charged pions, are sign selected 
to be positive by the horn system and allowed to decay in a
675 m long and 2 m diameter evacuated decay tunnel.   
The two-horn wide band system allows us to tune the mean 
 energy of the $\nu_\mu$ from about 3 to 18 GeV. Since the 
latest oscillation parameters from the Super Kamiokande 
atmospheric neutrino data 
are in the range of $\sim 3\times 10^{-3} eV^2$, the first run of 
MINOS will most likely be with the lowest energy beam to get  
as large an oscillation  effect as possible for the 735 km baseline. 
The expected $\nu_\mu$ flux 
from the NuMI beamline is shown 
in Figure \ref{flux}. The contamination from 
electron neutrinos will be less than $ 1\%$ with approximately 
the same spectrum.

\subsection{The MINOS detector}

\begin{figure}[htb]
\vspace{9pt}
\includegraphics[width=17pc]{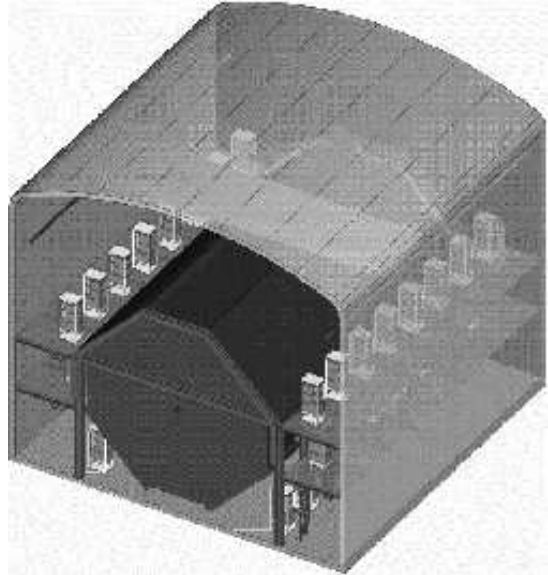}

\caption{Schematic of the MINOS far detector.
The octagonal area is the mainly iron detector mass;
the central hole accomodates a coil to generate the 
toroidal magnetic field. The racks on both sides of the 
detector house the multi-anode photomultipliers and the
front end electronics.  }
\label{det}
\end{figure}

The MINOS experiment has two detectors, the near detector and 
the far detector, to determine the neutrino flux close to the 
source as well as 735 km away.  
The near detector, with a mass of $\sim$ 1 kTon, will be placed 
1040 m downstream of the target, or about 300 m beyond the 
end of the decay tunnel, and about 100 m underground. It is 
designed to precisely measure neutrino beam characteristics 
such as the energy spectrum and the $\nu_e$ contamination. 
The near detector is constructed to be nearly identical to 
the far detector in terms of segmentation
so that the event reconstruction, 
background levels, and energy resolution can be studied with 
the unoscillated neutrino flux with high statistics.  
The neutrino interaction rate in the 
near detector is sufficient to provide  nearly continuous
beam monitoring.  For the final analysis we expect that 
interations in  the central 
region of 25 cm radius will be used.
Beam monitoring will also be provided by a set of 
ceramic pad ionization chambers 
placed in 3 caves after the end of the decay tunnel. These 
chambers were designed to provide stable, precise monitoring of
the secondary  muon flux as a function of  time as well as azimuthal 
position \cite{nim4}.

The far detector has a total mass of 5.4 kT. It is placed in 
a new cavern dug 713 m underground in the Soudan mine in 
northern Minnesota, about 735 km away from the primary target
at Fermilab. 
The far detector is  made out of two super-modules, each an
8m-diameter octagonal toroid composed of 243 
layers. Each layer is made of a 2.54 cm-thick steel plane and 
1 cm-thick and 4.1 cm-wide scintillator strips grouped in 20- or 
28-strip wide light-tight modules. The scintillator strips 
are made in an industrial extrusion process using 
Styron 663W polystryrene, manufactured by the DOW chemical company,
doped with 1\% PPO and 0.03\% POPOP. A 0.25 mm-thick reflective
layer, made by adding 12.5\% $\rm{TiO}_2$ to polystyrene, is 
co-extruded with the scintillator strips. A 1.4 mm-wide and 2 mm-deep 
groove in the center of the 4.1 cm-side is also made during the 
extrusion process. A 1.2 mm diameter wave length shifting (WLS) fiber
is embedded in the groove during the assembly of the 
scintillator modules. The J-type Y11 multiclad PMMA, non-S WLS fiber 
made by Kuraray and doped with the 
K27 dye at 175 ppm (with maximum intensity emission at 520 nm) is
used. The fibers are optically coupled to the scintillator 
strips with epoxy. The WLS fibers are read out from both ends. 
They are grouped (multiplexed) inside a light-tight box  into 
sets of 8 fibers from strips spaced more than 1 m apart in 
each plane. Each 8-fiber bundle is coupled, using a ``cookie'',
 to a single pixel of 
a 16-pixel  R5900-M16 Hamamatsu photomultiplier (PMT). 
 Thus each PMT reads out 128 fibers; 
one end of each scintillator plane needs 24 pixels. This arrangement
allows us to read out the entire MINOS far detector using only 
1452 PMTs. Since the event rate is small, unambiguous 
event recontruction can be achieved in software despite the 
multiplexing.  An important detector parameter is the 
photo-electron yield for a minimum ionizing particle (MIP) incident at 
right angle to the scintillator strip: the average yield,
measured using a radioactive source
 for each strip during assembly, is about 6 photo-electrons per MIP
 summed from
both sides.  The attenuation over the 8 meter length of the 
strip  is about a factor of 3.

The near detector is a 3.8 m by 4.8 m ``squeezed'' octagonal 
toroid made of solid 2.54 cm iron plates with scintillator 
strips of similar design as the far detector between the iron
plates.  The near detector has two longitudinal sections:
The first 96 planes are called the target or the hadron/shower 
calorimeter.
In these planes the area that will be exposed to the central 
part of the beam (about half of the tranverse area)
 has scintillator modules in every plane. Every 4th plane in 
the hadron/shower calorimeter has full area scintillator coverage.  
The second, muon tracking, section has 164 planes with every 5th gap 
instrumented  with full area scintillator coverage. 
The fibers are  read out individually (no multiplexing) 
from one end only by 215 64-anode Hamamatsu PMTs. The opposite ends 
of the fibers are reflectively coated to increase the average light 
yield to similar level as the far detector. 

Both detectors are magnetized using a coil through a hole in the
center of the planes to an average field of about 1.5 T (2 m away from the
coil). The front end electronics is different for the two detectors
because the event rate in the 8.1 $\mu$-sec neutrino pulse
 in the near detector is far  higher 
than the far detector.  In the far detector the read out electronics
is based on a VA chip from IDE and in the near detector it is 
based on the QIE chip designed at Fermilab.  Simulations show 
that these detectors have a resolution of $60\%/\sqrt{E}$ for 
hadronic showers and $25\%/\sqrt{E}$ for electromagnetic 
showers. Both detectors are being calibrated by cosmic rays 
and a light injection system. A test beam calibration module is 
being used to perform relative calibration between the near 
and far detectors of about 2\% and absolute calibration 
of about 5\%. 

The MINOS experiment is currently  under construction. 
Approximately 2/3 of the 
 detector has been assembled in the Soudan mine and is taking
cosmic ray data. The civil construction for the 
 NuMI beam line is finished and the technical items are now being 
assembled. Data with the neutrino beam is expected in December 2004. 

\section{The physics sensitivity}

A summary of the important parameters of the 
MINOS experiment is shown in Table \ref{table:1}.

\begin{table*}[htb]
\caption{Parameters of the MINOS experiment}
\label{table:1}
\newcommand{\m}{\hphantom{$-$}}
\newcommand{\cc}[1]{\multicolumn{1}{c}{#1}}
\renewcommand{\tabcolsep}{2pc} 
\renewcommand{\arraystretch}{1.2} 
\begin{tabular}{@{}ll}
\hline
Proton Energy & 120 GeV \\
Time Structure &  8.1 $\mu$ s every 1.9 sec \\
Intensity & $4\times 10^{13}$ protons/spill \\
Exposure &  $3.8\times 10^{20}$ protons/year \\
Beam & 1 m graphite target   2-horn pion focusing \\
     & LE 1-6 GeV $\nu_\mu$ with tail to 50 GeV \\
Baseline & 735.340 km  \\
Detector & 2.54 cm Iron, 1 cm scint. \\
	& Magnetized  Average field 1.5 T \\
Resolution	&  hadronic $60\%/\sqrt{E}$ \\
		&  electromagnetic $25\%/\sqrt{E}$ \\
Event rate   & Near det. 3 events/spill in target region \\
	& Far det.  300 $\nu_\mu$ CC events/kT/yr (no osc) \\
\hline
\end{tabular}\\[2pt]
\end{table*}

\subsection{Muon neutrino disappearance}

\begin{figure}[htb]
\vspace{9pt}
\includegraphics[width=17pc]{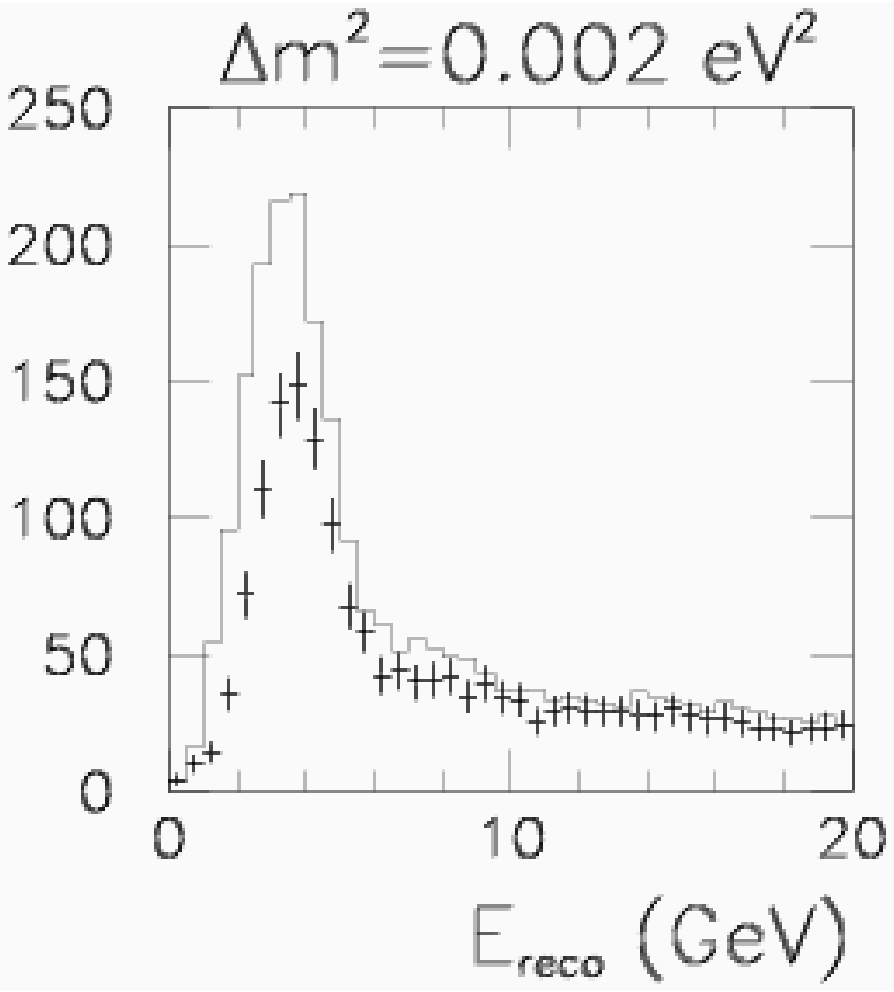}

\caption{Expected 
spectrum of reconstructed  muon neutrino energy for 
 $\Delta m^2=0.002~eV^2$ and $\sin^2 2 \theta = 0.9$.
 We assume 10kT-year of 
exposure in the low energy beam for this result.}
\label{edist1}
\end{figure}

\begin{figure}[htb]
\vspace{9pt}
\includegraphics[width=17pc]{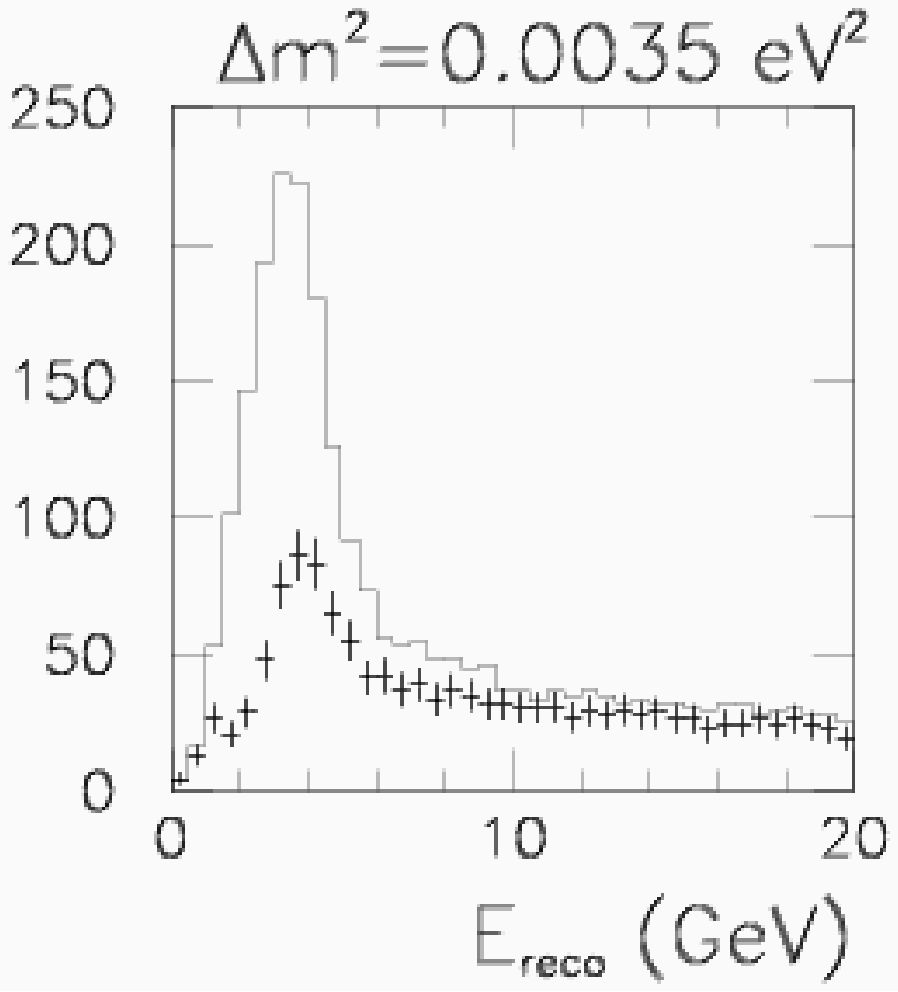}

\caption{Expected 
spectrum of reconstructed  muon neutrino energy for 
 $\Delta m^2=0.0035~eV^2$ and $\sin^2 2 \theta = 0.9$.
 We assume 10kT-year of 
exposure in the low energy beam for this result.}
\label{edist2}
\end{figure}

\begin{figure}[htb]
\vspace{9pt}
\includegraphics[width=20pc]{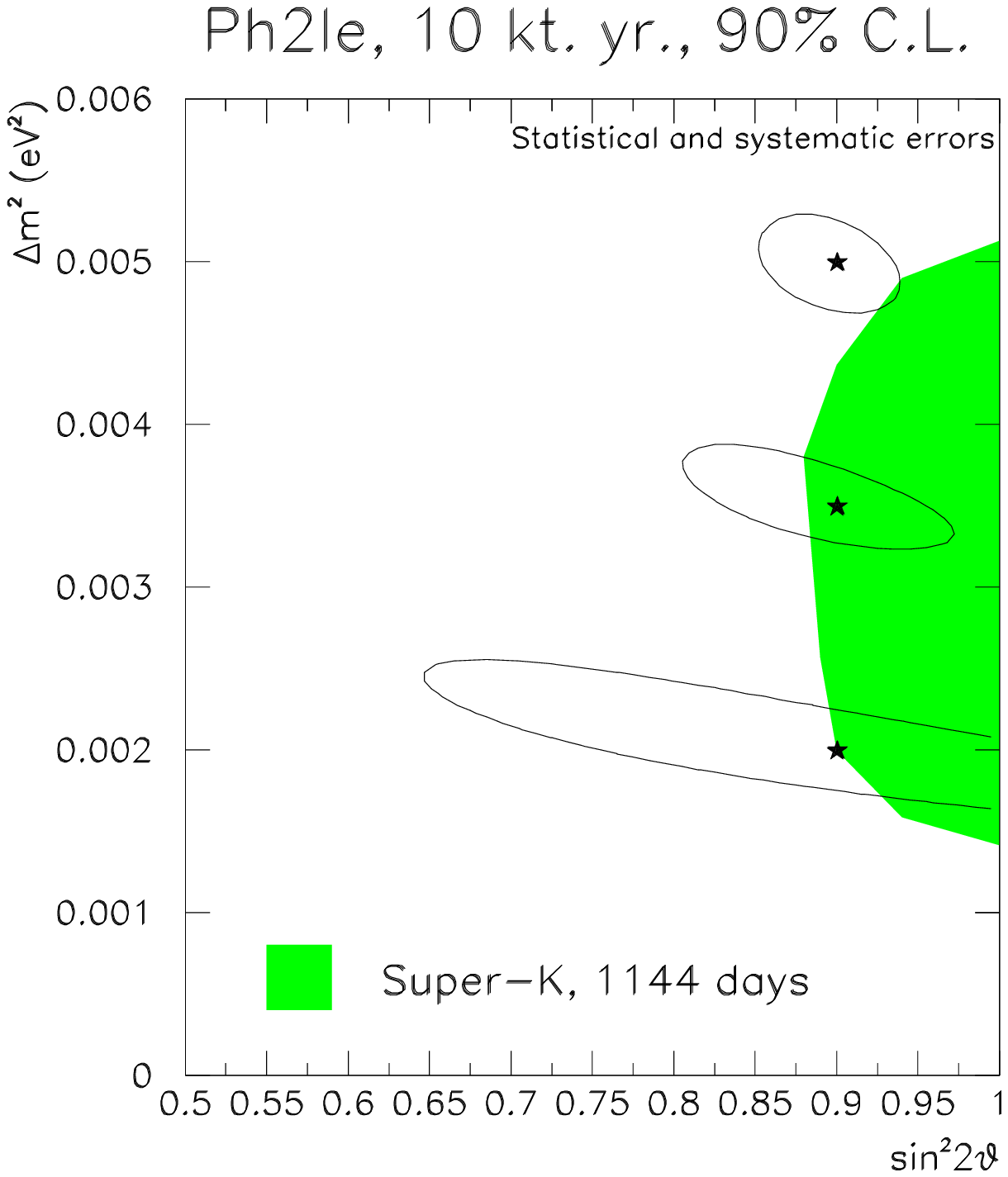}

\caption{Expected resolution from MINOS 
 from the disappearance signature for 
selected parameters (the stars). The 
favored region from Super Kamiokande data is also indicated.}
\label{ccsig}
\end{figure}

MINOS plans to detect oscillations by observing 
the absolute event rate as well as the energy distribution 
for charged current muon neutrino events. 
We will also obtain a confirming oscillation signature in 
the ratio of the number of charged  and neutral current events. 
 The total 
neutrino energy will be measured by  measuring the muon
momentum and range as well as the total energy of the hadron
shower. The resulting distribution
of reconstructed neutrino energy is shown in Figures 
\ref{edist1} and \ref{edist2}, for $\Delta m^2 = 0.002 ~eV^2$ 
and $\Delta m^2 = 0.0035 ~eV^2$, respectively; 
$\sin^2 2 \theta = 0.9$ for these
plots. To extract oscillation parameters it will be 
necessary to precisely predict the spectrum without oscillations 
at the far detector using the near detector data. 
Using known tolerances on the target/horn geometry 
as well as the differences in the near and far 
reconstrution efficiencies we expect to reduce the 
systematic error on the far spectrum to about 2\% in normalization 
and 2\% in bin-to-bin fluctuations. Using these systematic 
uncertainties we predict that MINOS will be able to measure 
the oscillation parameter $\Delta m^2$ 
 to $\pm 10\%$ at the best fit point of 
the Super-Kamiokande allowed region. This is shown in 
Figure \ref{ccsig}. 
The Super-Kamiokande best fit point is  now at $0.0025 ~eV^2$
\cite{rmp}. For this value, the minimum in the oscillation 
probability for MINOS will be at 1.5 GeV which is on the lower 
rising edge of the charged current spectrum (See the unoscillated 
histogram in Figure \ref{edist1}). 
Therefore,  one possibility under study for 
improved  performance is to increase the flux below 2 GeV by 
appropriate modifications to the target/horn geometry.

\subsection{Electron neutrino appearance}

\begin{figure}[htb]
\vspace{9pt}
\includegraphics[width=17pc]{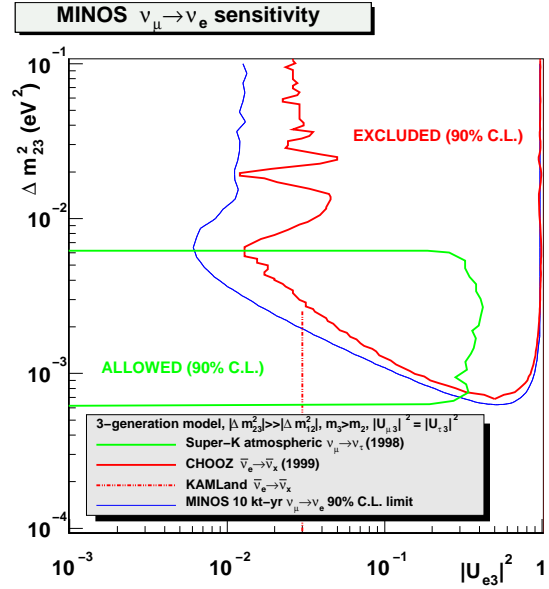}
\caption{Expected 90\% confidence level limit for 
$\nu_\mu \to \nu_e$ appearance if MINOS sees no 
excess of electron neutrino events with 10 kT-year of
exposure. }
\label{nue}
\end{figure}

The main oscillation channel in the case
of atmospheric oscillations is now thought to be 
$\nu_\mu \to \nu_\tau$\cite{sktau}.
Nevertheless, in the 3 neutrino formulation of oscillations 
it is natural to expect a small component of 
the disappearing $\nu_\mu$ to convert to $\nu_e$ with 
the same oscillation length as the atmospheric one.  
This will occur if the presently unknown 
neutrino mixing matrix element 
$U_{e3}$ is nonzero. The best limit on
$U_{e3}$ currently comes from the reactor experiment 
Chooz\cite{chooz} in which a search was made 
for disappearance of $\bar \nu_e$.  At the Super 
Kamiokande best fit
$\Delta m^2 = 0.0025 ~eV^2$ the Chooz limit implies
$|U_{e3}|^2 < 0.03$ at 90\% confidence level. 
The expected sensitivity of MINOS for 
$\nu_\mu \to \nu_e$ appearance is shown in 
Figure \ref{nue}. With 10 kT-year of exposure we expect to 
improve the Chooz limit by about a factor of 2. At 
$\Delta m^2 = 0.0025 ~eV^2$ and $|U_{e3}|^2 = 0.03$ 
we expect an excess of 24 electron event over a 
background of 40 events.
For Figure \ref{nue} we assumed that the 
background will be known to about 10\% by using the 
near detector data.  The sensitivity is limited by the 
neutral current background (28 events) out of which approximately 
half will be from high energy ($>10$ GeV) neutrinos. 
We are examining the possibility of reducing the background and 
improving the signal by 1) eliminating the high energy 
tail of the beam
by using a ``beamplug'', 2) improving the flux at low energies. 
A  more sensitive approach to  $\nu_\mu \to \nu_e$ 
 has been proposed to use the off-axis narrow band
 NuMI beam (See Figure \ref{flux})\cite{e889} 
and a new detector placed on the Earth's 
surface near Soudan\cite{offminos}. With an upgraded 
more intense main injector such an experiment could also 
be sensitive to CP violation.
Ultimately, for the best sensitivity to matter effects and
CP violation using a conventional neutrino 
beam one may have to 
go to much longer baselines \cite{billm,bnl69}.

\section{Conclusion}

The NuMI neutrino beamline and the MINOS near and 
far detectors are 
being  constructed. Approximately 2/3 of the 5.4 kT 
MINOS detector has been assembled in the Soudan mine; it is 
now gathering cosmic ray data.  The NuMI beamline will be finished in
2004 and data gathering with the neutrino beam will commence in 
December of 2004. 
With 10 kT-year of data with $3.8\times 10^{20}$ protons per year
we expect to confirm the 
oscillatory behaviour of muon neutrinos 
and measure $\Delta m^2$ with 10\% error.  We will also improve 
the limit on $\nu_\mu \to \nu_e$ beyond the limit of Chooz
or discover this appearance channel
if the parameter $|U_{e3}|^2$ is sufficiently large. 
The sensitivity to the appearance channel is limited by the
neutral current background, and we are examining various 
possibilities to reduce this background.


\begin{thebibliography}{9}
\bibitem{sk1} Y.~Fukuda et al., Phys. Rev. Lett. {\bf 81}, 1562 (1998) 

\bibitem{sktau} S. Fukuda et al., Phys. Rev. Lett. {\bf 85}, 3999 (2000)



\bibitem{sno1} Q.~R.~Ahmad et al., Phys. Rev. Lett. {\bf 87}, 071301 (2001) 

\bibitem{sno2} Q.~R.~Ahmad et al., Phys. Rev. Lett. {\bf 89}, 011301 (2002). 
 
\bibitem{sno3} Q.~R.~Ahmad et al., Phys. Rev. Lett. {\bf 89}, 011302 (2002).

\bibitem{lsnd}
C. Athanassopoulos et al., Phys. Rev. Lett. {\bf 77} 3082 (1996); 
 C. Athanassopoulos et al., Phys. Rev. Lett. {\bf 81} 1774 (1998) 

\bibitem{boon}
Booster Neutrino Experiment, Fermi National Laboratory, \\
http:/www-boone.fnal.gov/


                
\bibitem{mprop} Fermilab Proposal P875, February, 1995. The MINOS technical 
design report, Fermilab, NuMI-L-337, October, 1998. 

\bibitem{nim1} P.~Adamson et al.,  Nucl. Instrum. Meth. {\bf A492} 325-343
(2002)  

\bibitem{nim2} K.~Lang et al., Nucl. Instrum. Meth. {\bf A461}
571 (2001)  

\bibitem{nim3}  J. Hylen et al.,  FERMILAB-TM-2018, Sep 1997. 

\bibitem{nim4}  J. Mcdonald et al.,  Archive:physics/0205042, May 2002,  
To be published in  Nucl. Instrm. Meth.    


\bibitem{rmp} 
T. Kajita and Y. Totsuka, Rev. Mod. Phys. {\bf 73}, 85 (2001).


\bibitem{chooz} M.~Apollonio et al., Phys. Lett., {\bf B466}, 415 (1999). 

\bibitem{e889} Long Baseline Neutrino Oscillation Experiment, 
D. Beavis et al., 
Physics Design Report, BNL-52459. April 1995. 

\bibitem{offminos} Letter of Intent to build an Off-axis
 Detector to study $\nu_\mu  \to \nu_e$
  oscillations with the NuMI Neutrino Beam, D. Ayres et.al,
hep-ex/0210005

\bibitem{billm} William J. Marciano, 
arXiv: hep-phy/0108181,  22 Aug 2001.

\bibitem{bnl69}Report of the BNL Neutrino Working Group:
 Very Long Baseline Neutrino Oscillation 
      Experiment for Precise Determination of Oscillation 
      Parameters and Search for $\nu_\mu \to \nu_e$ Appearance 
      and CP Violation, 
M.~diwan, et al., BNL-69395, Oct 28, 2002. 
hep-ex/0211001. 


\end{thebibliography}
\end{document}